\documentclass{PoS}
\usepackage{slashed}
\usepackage{amsmath}
\usepackage{subfigure}

\title{Axial vector form factors in $D_s \rightarrow \phi$ semileptonic decays from lattice QCD}

\ShortTitle{Axial vector form factors in $D_s \rightarrow \phi$ semileptonic decays from lattice QCD}

\author{\speaker{Gordon Donald}, Christine Davies and Jonna Koponen, the HPQCD Collaboration\\
        SUPA, School of Physics and Astronomy, University of Glasgow, Glasgow G12 8QQ, UK\\
        E-mail: \email{g.donald@physics.gla.ac.uk}}

\abstract{We calculate axial vector and vector form factors for the semileptonic decay $D_s \rightarrow \phi$ using HISQ valence quarks on MILC ensembles with $2+1$ flavours of asqtad sea quarks. Using twisted boundary conditions to tune the quarks' momenta, we compute form factors at $q^2 =0$ on coarse lattices. We find $V(0) = 0.903(67) $, $A_1(0) = 0.603(20)$, $A_2(0) = 0.401(80)$ and $A_0(0) = 0.686(17)$, which we compare to experimental data and previous quenched lattice QCD calculations.}

\FullConference{ The XXIX International Symposium on Lattice Field Theory - Lattice 2011\\
July 10-16, 2011\\
Squaw Valley, Lake Tahoe, California}

\begin{document}
\section{Introduction}
The decay rate for the semileptonic meson decay $D_s \rightarrow \phi l \nu$ contains nonperturbative QCD form factors, about which we obtain information from lattice QCD.
The quark level transition is $c \rightarrow s$ and the decay is from a pseudoscalar meson to a vector.

Labelling the pseudoscalar as $D_s$ with momentum $p$ and the vector meson as $\phi$ with momentum $p'$ and polarisation vector $\varepsilon$, the general form for a pseudoscalar to vector matrix element is \cite{richmanburchat}:
\begin{align} \langle \phi(p', \varepsilon) | V^\mu - A^\mu | D_s (p) \rangle & = \frac{2i\epsilon_{\mu\nu\alpha\beta}}{m_{D_s}+m_{\phi}} \varepsilon^\nu p_{D_s}^\alpha p_\phi^\beta V(q^2) \label{matrixelement}
\\ & - (m_{D_s}+m_\phi)\varepsilon^\mu A_1(q^2) \nonumber 
\\ & + \frac{\varepsilon \cdot q}{m_{D_s}+m_\phi}(p+p')^\mu A_2(q^2) \nonumber
\\ & + 2m_\phi \frac{\varepsilon \cdot q}{q^2} q^\mu A_3(q^2) \nonumber
\\ & - 2m_\phi \frac{\varepsilon \cdot q}{q^2} q^\mu A_0(q^2) \nonumber
\end{align}
where we have the relations

\begin{equation} A_3(q^2) = \frac{m_{D_s}+m_\phi}{2m_\phi} A_1(q^2) - \frac{m_{D_s}-m_\phi}{2m_\phi} A_2(q^2) \end{equation}
and
\begin{equation} A_3(0) = A_0(0). \label{a3=a0} \end{equation}
$V(q^2)$ is the vector form factor.
There are three independent axial vector form factors as $A_3(q^2)$ is just a combination of $A_1(q^2)$ and $A_2(q^2)$ and at $q^2 = 0$ there are only two.
Calculating the form factors at $q^2 = 0$ allows us to compare to experiment for the experimentally accessible $A_1(q^2)$, $A_2(q^2)$ and $V(q^2)$ form factors.
BaBar \cite{babar} quote results for the quantities $r_2 = \frac{A_2(0)}{A_1(0)}$ and $r_V = \frac{V(0)}{A_1(0)}$.

\section{Methods}
\subsection{Lattice setup}
To compute semileptonic form factors, we calculate strange quark propagators using a random wall source on a timeslice at $x$.
A strange quark propagates to the sink, $y$, and then is used to make an extended propagator in which a charm quark propagates back to the current insertion at $z$, which is summed over a timeslice.
At the current insertion, the extended propagator is joined to the strange propagator to make the 3 point function.

We use HISQ \cite{hisq} valence quarks, which means that suitable phases for staggered quarks have to be inserted at the source, sink and current insertion.
The gauge configurations used are from MILC ensembles \cite{milcasqtad}, which include 2+1 flavours of asqtad sea quarks.
Currently we have performed calculations on only one ensemble, which has $\beta = 6.76$ and $V=20^3 \times 64$, so $a \sim 0.12$fm.
In general, the 3 point correlation function is
\begin{equation} \langle \bar{\psi}(x) \gamma_\mu \psi(x') \mbox{ } \bar{\psi}(z) \gamma_5 \gamma_\nu \psi(z') \mbox{ } \bar{\psi}(y) \gamma_5 \psi(y') \rangle, \end{equation}
where we allow for point splitting at the source, sink and current insertion.



\subsection{Twisted boundary conditions}
\begin{figure}
\begin{center}
\subfigure[Dispersion relation.]{\label{dispersion}\includegraphics[angle=-90,width=7.5cm]{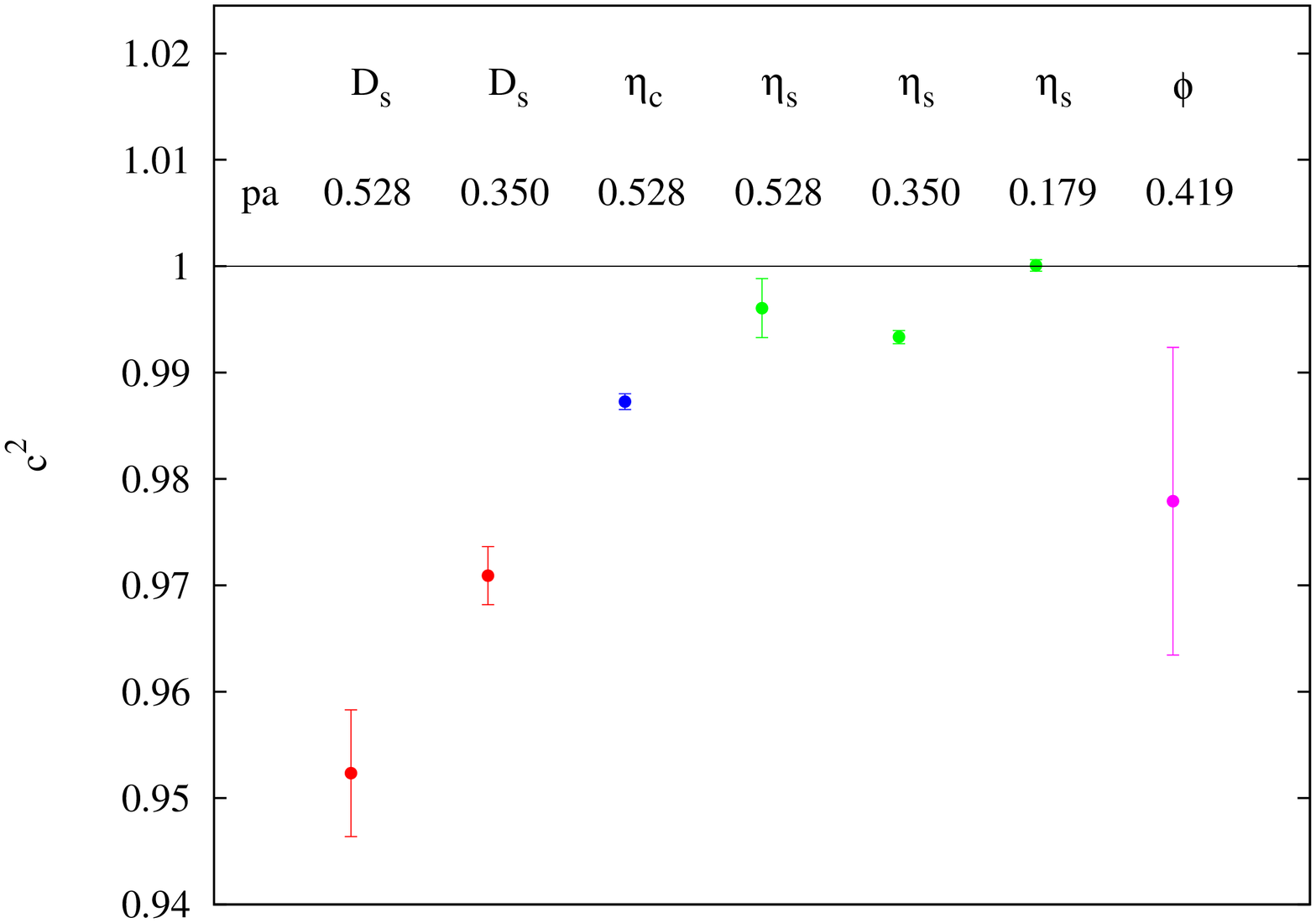}}
\subfigure[Ground state amplitude.]{\label{amplitudes}\includegraphics[angle=-90,width=7.5cm]{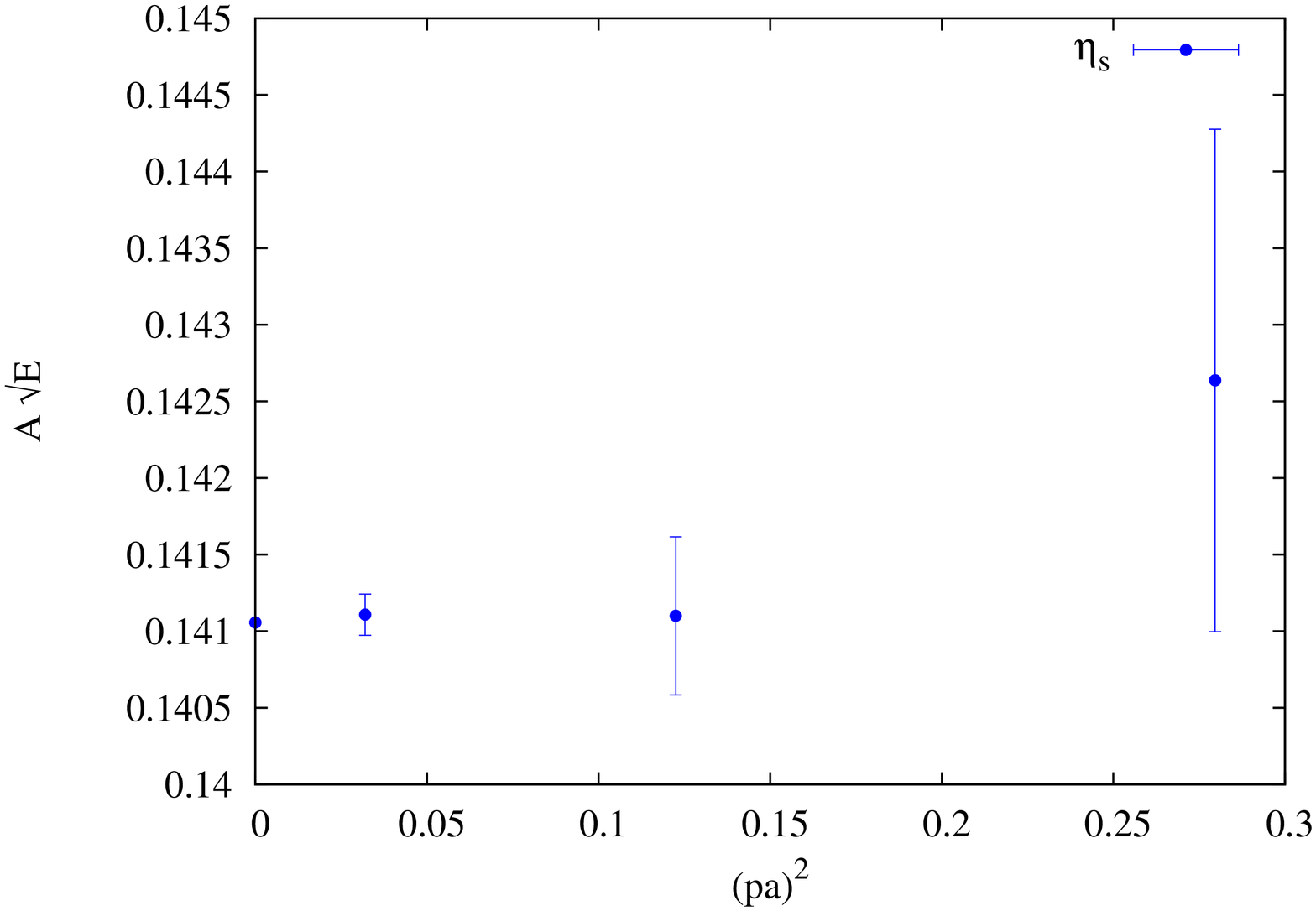}}
\end{center}
\caption{To check that introducing twisted boundary conditions gives us a physical momentum, we can calculate the speed of light for various mesons and momenta, $c^2 = \frac{E^2-m^2}{p^2}$ (Fig. \protect\ref{dispersion}) and look at the ground state's amplitude, $A$, as a function of meson energy. We expect $A \sqrt{E}$ to be constant (Fig. \protect\ref{amplitudes}).}
\label{twistedbcchecks}
\end{figure}

To calculate the form factor at $q^2 = 0$, we use a $\phi$ with non-zero momentum and a $D_s$ meson at rest.
We use twisted boundary conditions \cite{bedanque,etmctwist}, where we insert a tunable phase, $\theta$, at the boundary so $\psi(x+\hat{L}) = \psi(x)e^{i\theta}$.
In Figure \ref{twistedbcchecks}, we check that this introduces a physical momentum by looking at the dispersion relation and the amplitude of the ground state.



\subsection{$\phi$ 2 point correlators}
To extract matrix elements from 3 point correlators, we fit the 2 and 3 point functions simultaneously.
The 2 point functions give us the masses and decay constants of the $\phi$ and $D_s$ mesons in the 3 point amplitudes.
The $D_s$ mass is known to be accurately calculated with HISQ quarks \cite{precisionfds}.
We also extract $m_\phi$ and $f_\phi$, using either a taste-singlet 1-link or local vector operator.
Table \ref{phitable} gives the masses and decay constants of the 1-link and local $\phi$ mesons, which are in agreement.

\begin{table}
\begin{center}
\begin{tabular}{|c|c|c|c|}
\hline

 & 1-link & local & experiment \\ \hline

$m_\phi$ & 1086(4) MeV & 1083(4) MeV & 1019 MeV\\
$f_\phi$ & 248(4) MeV & 233(13) MeV & 227(4) MeV\\
 & $Z=1.095$ & $Z=0.97$ & PDG $\Gamma_{e^+e^-}$ \\
\hline
\end{tabular}
\end{center}
\caption{The masses and decay constants for our $\phi$ mesons calculated using the 1-link and local vector operators. We include experimental values from the PDG \cite{pdg} and note the lattice normalisation factors used for each calculation.}
\label{phitable}
\end{table}

The $\phi$ meson is not gold-plated, which means that it can decay to $K\bar{K}$ with a small width, an effect we do not fully include.
We have only done these calculations at one lattice spacing, so we cannot yet tell how much of the difference between the calculated lattice and continuum $\phi$ masses is caused by discretisation effects.
With results on finer lattices, the effects of the non-gold-platedness of the $\phi$ can be studied in more detail.
The normalisation factor for the 1-link vector operator is from Ref. \cite{jonnastalk} and a perturbative normalisation is used for the local vector \cite{perturbativenormalisation}.
We extract $f_\phi$ from the PDG results for $\Gamma_{e^+e^-}$ \cite{pdg}, using 
\begin{equation} \Gamma_{e^+e^-} = \frac{4\pi}{3} \alpha_{QED}^2 e_Q^2 \frac{f_\phi^2}{m_\phi}. \end{equation}

\subsection{Operator normalisation}
\subsubsection{1-link $A_\mu$}

To normalise the one link axial vector operator, we use simultaneous fits to
\begin{equation} \langle \bar{\psi}(x) \gamma_5 \psi(x) \mbox{ } \bar{\psi}(y) \gamma_5\gamma_\mu \psi(y+\hat{\mu}) \rangle 
\mbox{ and } \langle \bar{\psi}(x) \gamma_5 \psi(x) \mbox{ } \bar{\psi}(y) \gamma_5 \psi(y) \rangle, \end{equation}
making use of the facts that the operators $\bar{\psi}(x) \gamma_5 \psi(x)$ and $ \bar{\psi}(x) \gamma_5\gamma_\mu \psi(x+\hat{\mu})$ have the same taste and the pseudoscalar current is absolutely normalised.

\begin{figure}[h]
\begin{center}
\includegraphics[angle=-90,width=9cm]{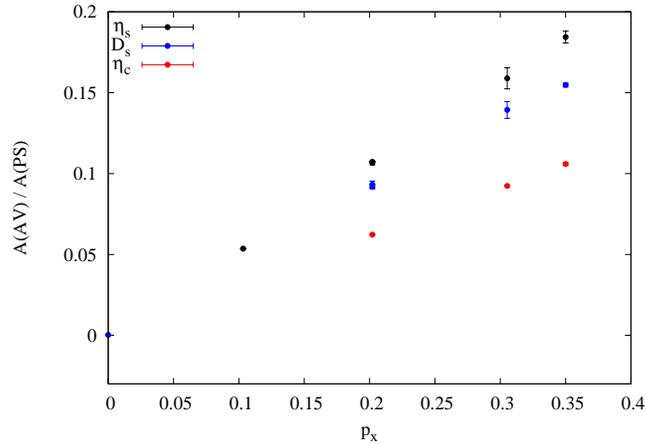}
\end{center}
\caption{The ratio of amplitudes for the pseudoscalar and axial vector operators goes as $\frac{p_\mu}{Z} \frac{m_1+m_2}{m_{P_0}^2}$, so the $Z$ factors can be extracted from the gradients of these lines.}
\label{zfactorsplot}
\end{figure}

The pseudoscalar and axial vector operators (in the continuum) can be related by
\begin{equation} p_\mu \langle 0 |A_\mu| P_0 \rangle = (m_1+m_2) \langle 0|P|P_0 \rangle. \end{equation}
The fit output gives the amplitude for both the pseudoscalar and the axial vector operators, which we obtain for several values of the meson momentum.
The ratio of amplitudes is a straight line against $p_\mu$ with a gradient of the $Z$ factor times the appropriate mass combinations.
In Figure \ref{zfactorsplot}, we plot this for various charm and strange pseudoscalar mesons.
For $cs$, the current we need for $D_s \rightarrow \phi$, we obtain the value $Z=1.065(5)$.

\subsubsection{Local $A_\mu$}
We can normalise the local $A_t$ component by creating non-Goldstone pseudoscalar mesons to compare to Goldstone mesons
\begin{equation} \langle \bar{\psi}(x) \gamma_5\gamma_t \psi(x) \mbox{ } \bar{\psi}(y) \gamma_5\gamma_t \psi(y) \rangle. \end{equation}
With HISQ, the masses are very close (for the $D_s$, the mass difference is 4.5 MeV) and the amplitudes, $a_{NG}$ and $a_G$ for the non-Goldstone and Goldstone mesons respectively, are related by
\begin{equation} Z E_{D_s} a_{NG} = (m_c+m_s) a_G, \end{equation}
where we have included the $Z$ factor for the local axial current.
For $cs$, we obtain $Z=1.036(4)$ for the local $A_\mu$ operator.
Note that because the masses are so close, we can also use non-Goldstone $D_s$ mesons in 3 point functions, which is useful when calculating the vector form factor, $V(q^2)$.

\section{$D_s \rightarrow \phi$ form factors}
\subsection{$A_1(q^2)$ form factor}
By setting up the $\phi$ polarisation and momentum such that $\varepsilon \cdot q = 0$, we can isolate the $A_1(q^2)$ form factor:
\begin{equation} \langle \phi(p', \varepsilon) | A^\mu | D_s (p) \rangle = (m_{D_s}+m_\phi)\varepsilon^\mu A_1(q^2). \end{equation}
We can use either the 1-link or local operator for the $\phi$ source, along with the appropriate axial vector current insertion.

The determinations of the $A_1(q^2)$ form factor using both the local and 1-link operators are plotted in Figure \ref{axialformfactors}, along with the other form factors for $D_s \rightarrow \phi$.




\subsection{$V(q^2)$ form factor} 
We can extract the vector form factor, $V(q^2)$, using a vector current insertion from Equation (\ref{matrixelement}) provided at least one of the $D_s$ and $\phi$ mesons is not at rest.
The 3 point function (with $D_s$ at rest) is
\begin{equation} \langle \bar{\psi}(x) \gamma_\mu \psi(x+\hat{\nu}) \mbox{ } \bar{\psi}(z) \gamma_\alpha \psi(z) \mbox{ } \bar{\psi}(y) \gamma_5 \gamma_t \psi(y) \rangle 
= \frac{2i\epsilon_{\mu\nu\alpha t}}{m_{D_s}+m_{\phi}} \varepsilon^\mu E_{D_s} p_\phi^\nu V(q^2), \end{equation}
where $\mu$, $\nu$ and $\alpha$ are the three spatial directions.
We use the non-Goldstone $D_s$ so need only include one link of point splitting to obtain an overall taste-singlet three point correlator for the pseudoscalar to vector transition with a local vector current insertion.

\begin{figure}[h]
\begin{center}
\includegraphics[angle=-90,width=9cm]{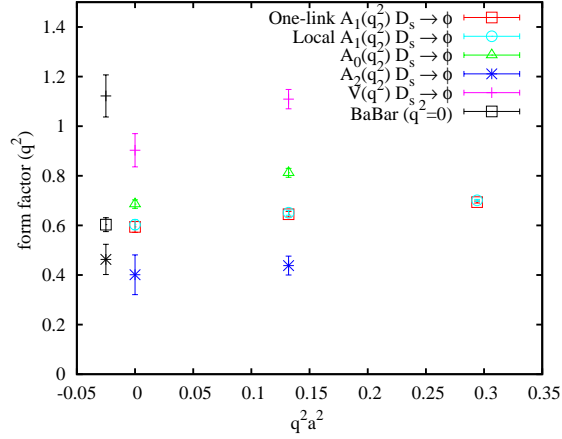}
\end{center}
\caption{The form factors for $D_s \rightarrow \phi$. Because of the kinematics, only $A_1(q^2)$ contributes at $q^2_{max}$ (where both the $D_s$ and $\phi$ mesons are at rest). All the form factors are calculated with $p_\phi a = 0.3$ and $p_\phi a = 0.4195$, where the latter corresponds to $q^2=0$. We also plot BaBar's values for the experimentally accessible $A_1(q^2)$, $A_2(q^2)$ and $V(q^2)$ form factors at $q^2=0$.}
\label{axialformfactors}
\end{figure}

We use a local vector normalisation of $Z=1.021(2)$ 
which is obtained \cite{jonnastalk} by comparing a non-Goldstone $D_s \rightarrow \eta_s$ decay through a local temporal vector current to the Goldstone $D_s \rightarrow \eta_s$ with a (normalised) 1-link vector current. 
$V(q^2)$ is plotted in Figure \ref{axialformfactors}.
\subsection{$A_0(q^2)$ and $A_2(q^2)$ form factors} 

The form factor $A_0(q^2)$ can be extracted using the pseudoscalar density:
\begin{align} (m_c + m_s) \langle \phi(p', \varepsilon) | P | D_s (p) \rangle & =  q_\mu \langle \phi(p', \varepsilon) | A^\mu | D_s (p) \rangle \\
 & =  2m_\phi \varepsilon \cdot q A_0(q^2). \nonumber 
\end{align}
The pseudoscalar form factor is not experimentally accessible, but is useful because it provides a route to obtaining $A_2(q^2)$. 

We can use Equation (\ref{a3=a0}) at $q^2 = 0$, $A_3(0) = A_0(0)$ to extract $A_2(0)$ from $A_0(0)$ and $A_1(0)$, although this does not hold at other values of $q^2$.
However we can calculate $\langle \phi(p', \varepsilon) | A_\mu | D_s (p) \rangle$ with $\varepsilon \cdot q \neq 0$ and use our determinations of $A_1(q^2)$ and $A_0(q^2)$ to find $A_2(q^2)$ with Equation (\ref{matrixelement}), provided $\frac{1}{q^2}$ is not too large.
\subsection{Comparison with experiment and previous calculations}
Table \ref{resultstable} compares our results at $q^2=0$ to BaBar's and those of previous quenched lattice QCD calculations. 
Our results for $A_2(0)$ and $r_2$ use our local $A_1(0)$ result, but the 1-link result is in close agreement.
We calculate $r_2$ directly from our results for $A_1(0)$ and $A_0(0)$ for better control of the errors.
\begin{table}
\begin{center}
\begin{tabular}{|c|c|c|c|}
\hline
 & HPQCD 2011 & BaBar 2008 \cite{babar} & UKQCD 2001 \cite{ukqcd} \\
\hline
$A_1(0)$ & $0.594 (22)$ & $0.607 (11) (19) (18)$ & $0.63 (2)$ \\
Local $A_1(0)$ & $0.603(20)$ & -- & -- \\
$V(0)$ & $0.903 (67)$ & $1.122(85)^*$ & $0.85 (4)$ \\
$A_0(0)$ & $0.686(17)$ & -- & $0.63(2)$ \\
$A_2(0)$ & $0.401(80)$ & $0.463(61)^*$ & $0.62(5)$ \\
$r_V$ & $1.52 (12)$ & $1.849 (60) (95)$ & $1.35(7)^*$ \\
$r_2$ & $0.62 (12)$ & $0.763(71)(65)$ & $0.98(8)^*$ \\
\hline
\end{tabular}
\end{center}
\caption{Table of form factors at $q^2=0$. We quote our determinations of $A_1(0)$ using 1-link and local operators separately. Entries marked with $^*$ have been extracted from the published results. BaBar's results include statistical and systematic errors. The third error quoted on their value of $A_1(0)$ is from theoretical input.}
\label{resultstable}
\end{table}

In Figure \ref{axialformfactors}, we plot the shape of the form factors as a function of $q^2$.
We expect the $q^2$ dependence to be relatively mild, and approximately the same, for all the axial vector form factors and steeper for the vector form factor.
BaBar assume the shape of the form factor and extract its value at $q^2=0$ from the total semileptonic rate.
Our results for $A_1(0)$ and $A_2(0)$ are in agreement with BaBar and our $V(0)$ result is a little lower.
In the previous quenched lattice QCD results, no difference was found between $A_1(0)$, $A_2(0)$ and $A_0(0)$. It is clear from Figure \ref{axialformfactors} that our results are different for these form factors at $q^2=0$.

\subsection{Future work}
%
To understand the discretisation effects better and also the importance of the non-gold-platedness of the $\phi$ mesons, we intend to repeat our calculations on MILC fine configurations, along with the determination of the $Z$ factors and semileptonic matrix elements.
We will also look at using lighter sea quark masses as, although neither the $\phi$ nor the $D_s$ contain light valence quarks, the non-gold-platedness of the $\phi$ may be an issue.

The $V(q^2)$ form factor is also important for radiative decays, where the photon is coupled by a vector operator, such as $J/\psi \rightarrow \eta_c \gamma$.
We will calculate this with the same method used here for $D_s \rightarrow \phi$.

We can extend our axial vector form factor calculations to heavier quarks.
Using HISQ for the heavy quark, we will extrapolate to the $b$ quark mass \cite{mbextrapolation} and calculate quantities such as $A_1(q_{max}^2)$ for $B \rightarrow D^*$.

\subsection*{Acknowledgements}
We are grateful to the MILC collaboration for the use of their configurations.
This work was funded by STFC.
The calculations were performed at the High Performance Computing Centre in Cambridge as part of the DiRAC facility, jointly funded by STFC, the Large Facilities Capital Fund of BIS, and the Universities of Cambridge and Glasgow.


\end{document}